\documentstyle[twoside,fleqn,espcrc2]{article}

\newcommand{\AmS}{{\protect\the\textfont2
  A\kern-.1667em\lower.5ex\hbox{M}\kern-.125emS}}

\title{Perturbative and Non-Perturbative Results for $N=2$ Heterotic
Strings\thanks{Talk presented by D. L\"ust}}

\author{Gabriel Lopes Cardoso\address{CERN, CH-1211 Geneva 23},
        Dieter L\"ust and Thomas Mohaupt\address{Institut f\"ur Physik,
        Humboldt Universit\"at Berlin,
        Invalidenstra{\ss}e 110, D-10115 Berlin}}

\def\caja{\mathsurround=0pt}
\def\eqalign#1{\,\vcenter{\openup1\jot \caja
       \ialign{\strut \hfil$\displaystyle{##}$&$
        \displaystyle{{}##}$\hfil\crcr#1\crcr}}\,}

\def\beq{\begin{equation}}
\def\eeq{\end{equation}}

\def\beq{\begin{equation}}
\def\eeq{\end{equation}}
\def\beqa{\begin{eqnarray}}
\def\eeqa{\end{eqnarray}}

\def\mxth{\mathsurround=0pt }
\def\xversim#1#2{\lower2.pt\vbox{\baselineskip0pt \lineskip-.5pt
x  \ialign{$\mxth#1\hfil##\hfil$\crcr#2\crcr\sim\crcr}}}

\def\beq{\begin{equation}}
\def\eeq{\end{equation}}
\def\beqa{\begin{eqnarray}}
\def\eeqa{\end{eqnarray}}

\newcommand{\be}{\begin{equation}}
\newcommand{\eq}{\end{equation}}

\newtheorem{satz}{Satz}[section]

\newtheorem{definition}{Definition}[section]

\newtheorem{bem}{Bemerkung}

\newtheorem{bsp}{Beispiel}

\begin{document}

\begin{abstract}
In this paper we summarize our recent work about perturbative and
non-perturbative effects in four-dimensional heterotic strings
with $N=2$ space-time supersymmetry.
\end{abstract}

\maketitle

\section{Introduction}

During the last year some major progress
has been obtained in the understanding  of the non-perturbative
dynamics of $N=2$ Yang-Mills theories as well as $N=2$ superstrings
in four dimensions.
In this paper we report on some work
concerning the computation of perturbative couplings for $N=2$ heterotic string
theories \cite{WKLL} as well as concerning the non-perturbative
monodromies \cite{trunc} in $N=2$ heterotic strings. We will show
that in the rigid limit the perturbative
and also the non-perturbative monodromies lead to
the monodromies discussed by Seiberg and Witten in \cite{SW1}.
Related  perturbative results  were independently obtained
in \cite{CAFP,AFGNT}. Our non-perturbative monodromies should
be compared \cite{comp} with the results \cite{KacVaf,many3,KKLMV,AntoPartou}
obtained from the string-string
duality between the heterotic and type  II strings with $N=2$
space-time supersymmetry.

\section{Classical results\label{classres}}

In this section we collect some results
about $N=2$ heterotic strings and the related classical prepotential;
we will in
particular work out the
relation between the enhanced gauge symmetries, the duality
symmetries and Weyl transformations. We will consider four-dimensional
heterotic vacua which are based on compactifications of
six-dimensional vacua on a two-torus $T_2$. The moduli of $T_2$
are commonly denoted by $T$ and $U$ where $U$ describes the
deformations of the complex structure, $U=(\sqrt G-iG_{12})/G_{11}$
($G_{ij}$ is the metric of $T_2$), while $T$ parametrizes the
deformations of the area and of the antisymmetric tensor, $T=2(\sqrt G+iB)$.
(Possibly other existing vector fields will not play any role in our
discussion.)
The scalar fields $T$ and $U$ are the spin-zero components of
two $U(1)$ $N=2$ vector supermultiplets.
All physical properties of the two-torus compactifications are
invariant under the group $SO(2,2,{\bf Z})$ of discrete target space
duality transformations. It contains the $T\leftrightarrow U$
exchange, with group element denoted by $\sigma$ and the
$PSL(2,{\bf Z})_T\times PSL(2,{\bf Z})_U$ dualities, which act on $T$
and $U$ as
\be
(T,U) \longrightarrow \left(
\frac{aT - ib}{icT + d}, \frac{a'U - ib'}{ic'U +d'} \right).
\eq
The classical monodromy group,
which is a true symmetry of the classical effective Lagrangian,
is generated by the elements $\sigma$,
$g_1$, $g_1$: $T\rightarrow 1/T$ and $g_2$, $g_2$: $T\rightarrow 1/(T-i)$.
The transformation $t$: $T\rightarrow
T+i$, which is of infinite order, corresponds to $t=g_2^{-1}g_1$.
Whereas $PSL(2,{\bf Z})_T$ is generated by $g_1$ and $g_2$,
the corresponding elements in $PSL(2,{\bf Z})_U$ are obtained
by conjugation with $\sigma$, i.e. $g_i'=\sigma^{-1}g_i\sigma$.

The $N=2$ heterotic string vacua
contain two further $U(1)$ vector fields, namely the graviphoton field,
which has no physical scalar partner, and the dilaton-axion field,
denoted by $S$. Thus the full Abelian gauge symmetry we consider is
given by $U(1)_L^2\times U(1)_R^2$.
At special lines in the $(T,U)$ moduli space, additional vector fields
become massless and the $U(1)_L^2$ becomes enlarged to a
non-Abelian gauge symmetry. Specifically, there are four
inequivalent lines in the moduli space where two charged gauge bosons
become massless: $U=T$,
$U=1/T$, $U=T-i$,
$U={T\over iT+1}$. The quantum numbers of the states that become massless
can be easily read of from the holomorphic mass formula \cite{OogVaf,FKLZ,CLM}
\begin{equation}
{\cal M}=m_2-im_1U+in_1T-n_2TU,
\label{HMF}
\end{equation}
where $n_i$, $m_i$ are the winding and momentum quantum numbers
associated with the $i$-th direction of the target space $T_2$.
At each of the four critical lines the $U(1)_L^2$ is extended to
$SU(2)_L\times U(1)_L$. Moreover, these lines intersect one another in two
inequivalent critical points (for a detailed discussion see \cite{CLM}).
At $(T,U)=(1,1)$ the first two lines
intersect. The four extra massless states extend the
gauge group to $SU(2)_L^2$. At $(T,U)=(\rho,\bar\rho)$ ($\rho=e^{i\pi/6}$)
the last three lines intersect. The six additional states extend the
gauge group to $SU(3)$. (In addition, the first and the third line
intersect at $(T,U)=(\infty,\infty)$, whereas the first and the last
line intersect at $(T,U) = (0,0)$.)

The Weyl groups of the enhanced gauge groups $SU(2)^2$ and $SU(3)$,
realized at $(T,U)=(1,1),(\rho,\bar\rho)$ respectively,
have the following action on $T$ and $U$ \cite{trunc}:
\be
\begin{array}{|l|l|l|} \hline
\mbox{Weyl Reflections} & T \rightarrow T' & U \rightarrow U' \\ \hline
w_1 & T \rightarrow U & U \rightarrow T \\
w_2 & T \rightarrow \frac{1}{U} & U \rightarrow \frac{1}{T}\\
w_1'&  T \rightarrow \frac{1}{U} & U \rightarrow \frac{1}{T}\\
w_2'&  T \rightarrow U+i & U \rightarrow T-i \\
w_0'& T \rightarrow \frac{U}{-iU+1} & U \rightarrow
\frac{T}{iT+1} \\ \hline
\end{array}
\label{WeylGroup}
\eq
$w_1$, $w_2$ are the Weyl reflections of $SU(2)_{(1)}\times SU(2)_{(2)}$,
whereas $w'_1$ and $w'_2$ are the fundamental Weyl reflections of the
enhanced $SU(3)$. For later reference we have also listed the $SU(3)$
Weyl reflection $w'_0={w'_2}^{-1} w'_1 w'_2$ at the hyperplane
perpendicular to the highest root of $SU(3)$. Note that $w_2=w'_1$.
All these Weyl transformations are target space modular transformations
and therefore elements of the monodromy group. All Weyl reflections
can be expressed in terms of the generators $g_1,g_2,\sigma$ and,
moreover, all Weyl reflections are conjugated to the mirror
symmetry $\sigma$ by some group element \cite{trunc}:
\be
w_1 = \sigma,\;\;\;
w_2 = w_1' = g_1 \sigma g_1 = g_1^{-1} \sigma g_1
\eq
\begin{equation}\eqalign{
w_2' &= t \sigma t^{-1} = (g_1^{-1} g_2)^{-1} \sigma (g_1^{-1}
g_2),\cr
w_0'& = w_2'^{-1} w_1' w_2'.\cr}
\label{ConjugatedToSigma}
\eq
As already mentioned the four critical lines are fixed under the corresponding
Weyl transformation. Thus it immediately follows that the numbers
of additional massless states agrees with the order of the fixed
point transformation at the critical line, points respectively \cite{CLM}.

Let us now express the moduli fields $T$ and $U$ in terms
of the field theory Higgs fields whose non-vanishing vacuum
expectation values spontaneously break the enlarged gauge
symmetries $SU(2)^2$, $SU(3)$ down to $U(1)^2$.
First, the Higgs field\footnote{Note that the
Higgs fields just correspond to the uniformizing variables
of modular functions at the critical points, lines respectively.}
 of $SU(2)_{(1)}$
is given by $a_1 \propto (T-U)$.
Taking the rigid field theory limit $\kappa^2={8\pi\over M_{\rm Planck}^2}
\rightarrow 0$ we will expand $T=T_0+\kappa \delta T$, $U=T_0+\kappa\delta U$.
Then, at the linearized level, the $SU(2)_{(1)}$ Higgs field is given
as $a_1  \propto (\delta T-\delta U) $.
Analogously, for the enhanced $SU(2)_{(2)}$ the Higgs field is $a_2
\propto (T-1/U)$.
Again, we expand as $T=T_0(1+\kappa\delta T)$, $U={1\over T_0}(1+\delta U)$
which leads to $a_2 \propto \delta T+\delta U$.
Finally, for the enhanced $SU(3)$ we obtain as Higgs fields
$a'_1 \propto \delta T+\delta U$,
$a_2' \propto \rho^2 \delta T + \rho^{-2} \delta U$,
where we have expanded as $T=\rho+\delta T$, $U=\rho^{-1}+\delta U$
(see section \ref{pertsec} for details).

The classical vector couplings are determined by the holomorphic prepotential
which is a homogeneous function of degree two of the fields $X^I$
($I=1,\dots , 3$). It is given by \cite{CAFP,WKLL,AFGNT}
\begin{equation}
F=i{X^1X^2X^3\over X^0}=- STU,
\label{classprep}
\end{equation}
where the physical vector fields are defined as $S=i{X^1\over X^0}$,
$T=-i{X^2\over X^0}$, $U=-i{X^3\over X^0}$ and the graviphoton corresponds
to $X^0$. As explained in \cite{CAFP,WKLL},
the period vector $(X^I,iF_I)$ ($F_I=
{\partial F\over X^I}$), that follows from the prepotential (\ref{classprep}),
does not lead to classical gauge couplings which all become small in the
limit of large $S$. Specifically, the gauge couplings which involve
the $U(1)_S$ gauge group are constant or even grow in the string weak
coupling limit $S\rightarrow\infty$. In order to choose a `physical'
period vector one has to replace $F_{\mu\nu}^S$ by its dual which is
weakly coupled in the large $S$ limit. This is achieved by the
following symplectic transformation $(X^I,iF_I)\rightarrow (P^I,iQ_I)$
where
$
P^1 = i F_1$, $Q_1 = i X^1$
and   $P^i = X^i$, $Q_i = F_i$    for  $i =0,2,3$.
In this new basis the classical period vector takes the form
\begin{equation}
\Omega^T=(1,TU,iT,iU,iSTU,iS,-SU,-ST),
\end{equation}
where $X^0=1$. One sees that in this new basis
all electric vector fields $P^I$ depend only on $T$ and $U$, whereas
the magnetic fields $Q_I$ are all proportional to $S$.

The basis $\Omega$ is also well adapted to discuss the action
of the target space duality transformations and, as particular
elements of the target space duality group, of the four inequivalent
Weyl reflections given in (\ref{WeylGroup}).
These transformations are given by
symplectic $Sp(8,{\bf Z})$ transformations, which act on the period
vector $\Omega$ as
\begin{equation}
\pmatrix{P^I\cr i Q_I\cr}\rightarrow \Gamma\pmatrix{
P^I\cr  i Q_I\cr}=\pmatrix{U&Z\cr W&V\cr}\pmatrix{
P^I\cr i Q_I\cr},
\label{symptr}
\end{equation}
where the $4\times 4$ sub-matrices $U,V,W,Z$ have to satisfy the symplectic
constraints
$
U^T V - W^T Z = V^T U - Z^T W = 1$,
$U^T W = W^T U$, $Z^T V = V^T Z$.
Invariance of the lagrangian implies that $W=Z=0$, $VU^T=1$. In case
that $Z=0$, $W\neq 0$ and hence $VU^T=1$ the action is invariant up
to shifts in the $\theta$-angles; this is just the situation  one
encounters at the one-loop level. The non-vanishing matrix $W$
corresponds to non-trivial one-loop monodromy due to logarithmic
singularities in the prepotential. (This will be the subject of section 3.)
Finally, if $Z\neq 0$ then the electric fields transform into magnetic fields;
these transformations are the non-perturbative monodromies due to
logarithmic singularities induced by monopoles, dyons or other
non-perturbative excitations (see section 4).

The classical action is completely invariant under the target space
duality transformations. Thus the classical monodromies have $W,Z=0$.
The matrices $U$ (and hence $V=U^{T,-1}=U^*$) are given by
\begin{equation}\eqalign{
U_{\sigma} &= \left( \begin{array}{cccc}
1&0&0&0\\
0&1&0&0\\
0&0&0&1\\
0&0&1&0\\
\end{array} \right),\cr
U_{g_1}& = \left( \begin{array}{cccc}
0&0&1&0\\
0&0&0&1\\
-1&0&0&0\\
0&-1&0&0\\
\end{array} \right),\cr
U_{g_2}& = \left( \begin{array}{cccc}
1&0&1&0\\
0&0&0&1\\
-1&0&0&0\\
0&-1&0&1\\
\end{array} \right) .\cr}
\end{equation}
At the classical level the $S$-field is invariant under these
transformations.
The corresponding symplectic matrices for the four inequivalent
Weyl reflections  then immediately follow from the previous equations (4) and
(5).

\section{Perturbative results \label{pertsec}}

Let us first review the main results about the one-loop perturbative
holomorphic prepotential which were derived in \cite{WKLL,AFGNT}.
Using simple power counting arguments it is clear that the one-loop
prepotential must be independent of the dilaton field $S$. The same
kind of arguments actually imply that there are no higher loop corrections
to the prepotential in perturbation theory. Thus the perturbative, i.e.
one loop prepotential takes the form
\begin{equation}\eqalign{
F &= F^{(Tree)}(X) + F^{(1-loop)}(X)\cr
&= i \frac{X^1 X^2 X^3}{X^0} + (X^0)^2 f(T,U) \cr &= - STU + f(T,U).\cr}
\end{equation}
Since the target space duality transformations are known to be
a symmetry in each order of perturbation theory, the tree level plus
one-loop effective action must be invariant under these transformations, where
however one has to allow for discrete shifts in the various $\theta$
angles
due to monodromies around semi-classical
singularities in the moduli space where massive string modes become
massless.
Instead of the classical transformation rules,
in the quantum theory, $(P^I,i Q_I)$ transform according to
\begin{equation}
P^I\ \rightarrow\    U{}^I_{\,J}\, P^J,\qquad
i Q_I\ \rightarrow\    V_I{}^J \,i Q_J\ +\   W_{IJ}\, P^J\
\label{symptrans}
\end{equation}
where
\begin{equation}
 V = ( U^{\rm T})^{-1},\quad
 W =  V \Lambda\,,\quad \Lambda=\Lambda^{\rm T}
\end{equation}
and $ U$ belongs to  $SO(2,2,{\bf Z})$.
Classically, $\Lambda=0$, but in the quantum theory, $\Lambda$
is a real symmetric matrix, which should be integer
valued in some basis.

Besides the target space duality symmetries, the effective action is
also invariant, up to discrete shifts in the $\theta$-angles,
under discrete shifts in the $S$-field, $D$: $S\rightarrow S-i$.
Thus the full perturbative monodromies contain the following $Sp(8,{\bf Z})$
transformation:
\begin{equation}\eqalign{
V_{S} &= U_S = \left( \begin{array}{cccc}
1&0&0&0\\
0&1&0&0\\
0&0&1&0\\
0&0&0&1\\
\end{array} \right),\cr
W_S &= \left( \begin{array}{cccc}
0&1&0&0\\
1&0&0&0\\
0&0&0&1\\
0&0&1&0\\
\end{array} \right),\;\;\; Z_S = 0.\cr}
\label{sdualpert}
\end{equation}

Invariance of the one-loop action up to discrete $\theta$-shifts then implies
that
\be
F^{(1-loop)} (X) \longrightarrow F^{(1-loop)} (X) - \frac{i}{2}
\Lambda_{IJ} P^I P^J
\eq
This reads in special coordinates like
\be
f(T,U) \rightarrow (icT + d)^{-2} (f(T,U) + \Psi(T,U))
\eq
for an arbitrary $PSL(2,{\bf Z})_T$ transformation.
$\Psi(T,U)$ is a quadratic polynomial in $T$ and $U$.

As explained in \cite{WKLL,AFGNT}
the dilaton is not any longer invariant under
the target space duality transformations at the one-loop level.
Indeed, the relations
(\ref{symptrans}) imply
\be
S \longrightarrow S + \frac{V_1^{\;J} (F^{(1-loop)}_J -i
\Lambda_{JK} P^K )}{U^{0}_{\;I} P^I}.
\label{spert}
\eq

Near the singular lines the one-loop prepotential exhibits
logarithmic singularities and is therefore not a singlevalued function
when transporting the moduli fields around the singular lines.
For example around the singular $SU(2)_{(1)}$ line $T=U \neq 1,\rho$
the function $f$
must have the
following form
\begin{equation}
f(T,U)={1\over \pi}(T-U)^2\log(T-U)+\Delta(T,U),
\end{equation}
where $\Delta(T,U)$ is finite and single valued at $T=U \neq 1,\rho$.
At the remaining three critical  lines $f(T,U)$ takes an analogous form.
Moreover at the intersection points the residue of the
singularity must change in agreement with the number of states
which become massless at these critical points.  (These residues are
of course just given by the $N=2$ pure Yang-Mills $\beta$-functions
for $SU(2)$, $SU(2)^2$ and $SU(3)$ (there are no massless additional flavors
at the points of enhanced symmetries).)
Specifically at the point $(T,U)=(1,1)$ the
prepotential takes the form
\begin{equation}
f(T,U=1)={1\over \pi}(T-1)\log(T-1)^2+\Delta'(T)
\end{equation}
and around $(T,U)=(\rho,\bar\rho)$
\begin{equation}
f(T,U={\bar \rho})={1\over \pi}(T-\rho)\log(T-\rho)^3+\Delta''(T),
\end{equation}
where $\Delta'(T)$, $\Delta''(T)$ are finite at $T=1$, $T=\rho$ respectively.
$f(T,U)$ is not a true modular form, but has non-trivial monodromy
properties; remarkably, a
closed, analytic expression for
$f(T,U)$ in terms of the trilogarithm function was recently derived in
\cite{Moore}.
However the third derivative transforms nicely
under target space duality transformations, and using the informations
about the order of poles and zeroes one can uniquely determine
\cite{WKLL,AFGNT}
\[
\partial^3_T f(T,U) \propto  \frac{+1}{2 \pi}
\frac{E_4(iT) E_4(iU) E_6(iU) \eta^{-24}(iU)}{j(iT) - j(iU)}
\]
\be
\partial^3_U f(T,U) \propto \frac{-1}{2 \pi}
\frac{E_4(iT) E_6(iU) \eta^{-24}(iT) E_4(iU)}{j(iT) - j(iU)}
\label{ThirdDerivative}
\eq
This result has recently prooved \cite{KacVaf,many3}
to be important to support the
hypotheses
that the quantum vector moduli space of the $N=2$ heterotic string
is given  by the tree level vector moduli space of a dual type II,
$N=2$ string, compactified on a suitably choosen Calabi-Yau space.
In addition to eq.(\ref{ThirdDerivative}) one can also deduce that
\cite{CLM,WKLL}
\be
\partial_T \partial_U f =- \frac{2}{\pi}
\log(j(iT) - j(iU)) + h(T,U),
\eq
where $h(T,U)$ is finite in the entire modular region.
$\partial_T\partial_U f$ provides the gauge threshold
corrections for the $U(1)_L^2$ gauge groups. It
has precisely the right property that the coefficient
of the logarthmic singularity is proportional to the number
of generically massive states that become massless.
The function $h(T,U)$ enters the defintion of a dilaton field
which is invariant under target space duality transformations \cite{WKLL}:
$S^{\rm inv}=S-{1\over 2}h(T,U)$.

\subsection{Perturbative $SU(2)_{(1)}$ monodromies}

Let us now consider the element $\sigma$
which corresponds to the Weyl reflection in the first enhanced $SU(2)_{(1)}$
Under the
mirror transformation $\sigma$,
$T \leftrightarrow U, T-U \rightarrow e^{-i \pi}
(T-U)$, and the
$P$ transform classically and perturbatively as
\be
P^0 \rightarrow P^0,\;\;\;
P^1 \rightarrow P^1,\;\;\;
P^2 \rightarrow P^3,\;\;\;
P^3 \rightarrow P^2
\eq
The one--loop correction $f(T,U)$ transforms as\footnote{
Note that one can always add polynomials of quadratic order
in the moduli to a given $f(T,U)$ \cite{AFGNT}. This results
in the conjugation of the monodromy matrices.  Hence, all
the monodromy matrices given in the following are unique up to
conjugation.}
\beqa
f(T,U) &\rightarrow& f(U,T) = f(T,U) -i(T-U)^2 \nonumber\\
f_T(U,T) &=& f_U(T,U) + 2i (T-U),\nonumber\\
f_U(U,T) &= &f_T(T,U) - 2i (T-U)
\eeqa
The $f$ function must then have the following form for
$T \rightarrow U$
\be
f(T,U) =  \frac{1}{\pi} (T-U)^2 \log(T-U) + \Delta(T,U)
\label{fisu2}
\eq
with derivatives
\beqa
f_T(T,U) &= & \frac{2}{\pi} (T-U) \log(T-U)
\nonumber\\ &+& \frac{1}{\pi} (T-U) +
\Delta_T
\nonumber\\
f_U(T,U) &=&  -\frac{2}{\pi} (T-U) \log(T-U) \nonumber\\ &-& \frac{1}{\pi}
(T-U)
+ \Delta_U
\label{fsu2tu}
\eeqa
$\Delta(T,U)$ has the property that it is finite as $T \rightarrow U \neq
1,\rho$ and that, under mirror symmetry $T \leftrightarrow U$,
$\Delta_T \leftrightarrow \Delta_U$.
The 1-loop corrected $Q_2$ and $Q_3$ are thus given by
\beqa
Q_2 &= &i S U
- \frac{2i}{\pi} (T-U) \log(T-U)
\nonumber\\ &-& \frac{i}{\pi} (T-U) -i \Delta_T \nonumber\\
Q_3& =& i S T
+ \frac{2i}{\pi} (T-U) \log(T-U) \nonumber\\ &+ &\frac{i}{\pi} (T-U)
-i \Delta_U
\label{q2q3}
\eeqa

It follows from (\ref{spert}) that,
under mirror symmetry $T \leftrightarrow U$,
the dilaton $S$ transforms as
\be
S \rightarrow S + i
\label{stransf}
\eq
Then, it follows that
perturbatively
\beqa
\left( \begin{array}{c}
Q_2\\  Q_3\\
\end{array} \right)
\rightarrow
\left( \begin{array}{c}
Q_3\\  Q_2\\
\end{array} \right) +
\left( \begin{array}{cc}
1 & -2\\
-2 & 1 \\
\end{array} \right)
\left( \begin{array}{c}
T\\  U\\
\end{array} \right)
\label{su2mono}
\eeqa
Thus,
the section $\Omega$ transforms perturbatively as
$\Omega \rightarrow \Gamma_{\infty}^{w_1} \Omega$, where
\beqa
\Gamma_{\infty}^{w_1} &=&
\left( \begin{array}{cc}
U & 0 \\ U \Lambda & U\\
\end{array} \right)  \;\;,\;\;
U=\left( \begin{array}{cc}
I & 0 \\ 0 &  \eta \\
\end{array} \right),\nonumber\\
\Lambda &=&-
\left( \begin{array}{cc}
\eta & 0 \\
0 & {\cal C}\\
\end{array} \right) \;\;,\;\;
\eta = \left( \begin{array}{cc}
0& 1 \\
1 & 0\\
\end{array} \right),\nonumber\\
{\cal C}& = &\left( \begin{array}{cc}
2& -1 \\
-1 & 2\\
\end{array} \right)
\nonumber\\
\label{perttrans}
\eeqa

\subsection{Truncation to the rigid case of Seiberg/Witten \label{secseiwit}}

In order to truncate the monodromies of local $N=2$ supergravity to
the case of rigid $N=2$ super Yang-Mills one has to take the limit
$M_{pl}\rightarrow\infty$ in an appropriate way. Specifically,
one has to freeze, i.e. get rid off the dilaton and graviphoton degrees
of freedom. This amounts to truncate the $8\times 8$, $Sp(8)$ monodromy
matrices
(in case of two Higgs fields) of
local supergravity to $4\times4$, $Sp(4)$ monodromy matrices of
rigid supersymmetry. These rigid $Sp(4)$ matrices act within the
four-dimensional subspace spanned by
the $(P^2,P^3,iQ_2,iQ_3)$ which is related to the two Higgs fields $a_1$,
$a_2$ and their duals $a_{D1}$, $a_{D2}$.
However this truncation procedure requires some care since
in the string case the
dilaton as well as the graviphoton are in general not invariant under
the Weyl transformations. It follows that the truncated $Sp(4)$ matrices
are not simply given by the $4\times 4$ submatrices of the local monodromies
as we will show in the following.\footnote{As an alternative \cite{AntoPartou}
one can first introduce
a compensating
shift for the dilaton, $S \rightarrow S - i$, which is generated
by $\Gamma_S$ in (\ref{sdualpert}).
Then the rigid monodromies are immediately given
by the four-dimensional sub-matrices of the
local monodromy matrices. This corresponds to a different, but
equivalent freezing in of the dilaton.
We will discuss these issues in a different paper \cite{comp}.}

In order to truncate \cite{trunc} the perturbative $SU(2)_{(1)}$ monodromy
$\Gamma_{\infty}^{w_1}$
to the rigid one of Seiberg/Witten \cite{SW1}, we will  expand
\beqa
T = T_0 + \kappa \delta T,\quad
U= T_0 + \kappa \delta U
\eeqa
Here we have expanded the moduli fields $T$ and $U$ around the same
vev $T_0 \neq 1,\rho$.  Both $ \delta T$ and $\delta U$ denote
fluctuating fields of mass dimension one.  We will also freeze in the
dilaton field to a
large vev, that is we will set $S = \langle S \rangle \rightarrow
\infty$.
Then, the $Q_2$ and $Q_3$ given in (\ref{q2q3}) can be expanded
as
\beqa
Q_2 &=&  i \langle S \rangle T_0 + \kappa \tilde{Q}_2 \;\;\;,\;\;\;
Q_3 = i \langle S \rangle T_0 + \kappa \tilde{Q}_3 \nonumber\\
\tilde{Q}_2 &=& i \langle S \rangle \delta U
- \frac{2i}{\pi} (\delta T- \delta U)
\log\kappa^2 ( \delta T - \delta U)
\nonumber\\ &- &\frac{i}{\pi} (\delta T- \delta U)
-i \Delta_T(\delta T,\delta U) \nonumber\\
\tilde{Q}_3 &=& i \langle S \rangle \delta T
+ \frac{2i}{\pi} (\delta T- \delta U)
\log\kappa^2 ( \delta T - \delta U) \nonumber\\ &+ &\frac{i}{\pi}
 (\delta T- \delta U)
-i \Delta_U(\delta T,\delta U) \nonumber\\
\label{truncq2q3}
\eeqa
Next, one has to specify how mirror symmetry is to act on
the vev's $T_0$ and $\langle S \rangle$ as well as on $\delta T$ and
$\delta U$.  We will take that under mirror symmetry
\beqa
T_0 \rightarrow T_0 \;\;,\;\; \delta T \leftrightarrow \delta U \;\;,\;\;
\langle S \rangle \rightarrow
\langle S \rangle
\label{tSTU}
\eeqa
Note that we have taken $\langle S \rangle$ to be invariant under mirror
symmetry.  This is an important difference to (\ref{stransf}).
Using (\ref{tSTU}) and that $\delta T - \delta U \rightarrow
e^{-i \pi} (\delta T - \delta U)$, it follows that the truncated
quantitities $\tilde{Q}_2$ and $\tilde{Q}_3$ transform as
follows under mirror symmetry
\beqa
\left( \begin{array}{c}
\tilde{Q}_2\\  \tilde{Q}_3\\
\end{array} \right)
\rightarrow
\left( \begin{array}{c}
\tilde{Q}_3\\  \tilde{Q}_2\\
\end{array} \right) +
\left( \begin{array}{cc}
2 & -2\\
-2 & 2 \\
\end{array} \right)
\left( \begin{array}{c}
\delta T\\ \delta  U\\
\end{array} \right)
\label{truncsu2mono}
\eeqa
Defing a truncated section $\tilde{\Omega}^T= (\tilde{P}^2,\tilde{P}^3,
i\tilde{Q}_2,i\tilde{Q}_3)=(i \delta T,i \delta U,
i\tilde{Q}_2,i\tilde{Q}_3)$, it follows that $\tilde{\Omega}$ transforms
as $\tilde{\Omega} \rightarrow \tilde{\Gamma}_{\infty}^{w_1} \tilde{\Omega}$
under mirror symmetry (\ref{tSTU}) where
\beqa
\tilde{\Gamma}_{\infty}^{w_1}& = &
\left( \begin{array}{cc}
\tilde{U} & 0 \\ \tilde{U} \tilde{\Lambda} & \tilde{U}\\
\end{array} \right)  \;\;,\;\;
\tilde{U}= \eta,\nonumber\\
\eta &= &\left( \begin{array}{cc}
0& 1 \\
1 & 0\\
\end{array} \right)\;\;,\;\;
\tilde{\Lambda} = \left( \begin{array}{cc}
-2& 2 \\
2 & -2\\
\end{array} \right)
\label{truncmonotrans}
\eeqa
Note that, because of the invariance of $\langle S \rangle$ under mirror
symmetry, $\tilde{\Lambda} \neq - {\cal C}$, contrary to what one
would have gotten by performing a naive truncation of (\ref{perttrans})
consisting in keeping only rows and columns associated with
$(P^2,P^3,iQ_2,iQ_3)$.

Finally, in order to compare the truncated $SU(2)$ monodromy
(\ref{truncmonotrans}) with the perturbative $SU(2)$ monodromy
of Seiberg/Witten \cite{SW1}, one has to perform a change of basis from
moduli fields to Higgs fields, as follows
\beqa
\left( \begin{array}{c}
a \\
a_D\\
\end{array} \right)
&=&
M \tilde{\Omega} \;\;,\;\;
M =
\left( \begin{array}{cc}
m &  \\
 & m^* \\
\end{array} \right),\nonumber\\
m&= &\frac{\gamma}{\sqrt{2}}
\left( \begin{array}{cc}
1& -1 \\
1& 1\\
\end{array} \right)
\label{higgsbasis}
\eeqa
where $\gamma$ denotes a constant to be fixed below.  Then, the
perturbative $SU(2)$ monodromy in the Higgs basis is given by
\beqa
\tilde{\Gamma}^{Higgs}_{\infty}& = &M \tilde{\Gamma}_{\infty}^{w_1} M^{-1}
\nonumber\\ &=&
\left( \begin{array}{cc}
m \tilde{U} m^{-1}  &   0  \\
m^* \tilde{U} \tilde{\Lambda} m^{-1} &
m^* \tilde{U} m^T \\
\end{array} \right)
\eeqa
which is computed to be
\beqa
\tilde{\Gamma}^{Higgs,w_1}_{\infty} =
\left( \begin{array}{cccc}
-1 & & & \\
& 1 & & \\
\frac{4}{\gamma^2} & 0 & -1 & 0 \\
& & & 1\\
\end{array} \right)
\label{higgsmono}
\eeqa
Note that (\ref{higgsmono}) indeed correctly shows that, under the
Weyl reflection in the first $SU(2)$, the second $SU(2)$ is left untouched.
The fact that (\ref{higgsmono}) reproduces this behaviour
can be easily traced back to the fact that we have assumed that $
\langle S \rangle$ stays invariant under the mirror transformation
$\delta T \leftrightarrow \delta U$.  Finally, comparing with the
perturbative $SU(2)$ monodromy of Seiberg/Witten \cite{SW1} yields that
$\gamma^2 = 2$, whereas comparision with the perturbative $SU(2)$
monodromy of Klemm et al \cite{KLTY,KLT} gives that $\gamma^2 = 1$.

The discussion about the other Weyl twist is completely analogeous
to $w_1$ and can be found in \cite{trunc}.

\section{Non-perturbative monodromies}



In order to obtain some information about non-perturbative
monodromies in $N=2$ heterotic string compactifications, we will follow
Seiberg/Witten's strategy in the rigid case \cite{SW1}
and try to decompose the
perturbative monodromy
matrices $\Gamma_{\infty}$ into
$\Gamma_{\infty} = \Gamma_M \Gamma_D$ with $\Gamma_M$ ($\Gamma_D$)
possessing monopole like (dyonic) fixed points. Thus each semi-classical
singular line will split into two non-perturbative singular lines where
magnetic monopoles or dyons respectively become massless. In doing so
we will
work in the limit of large dilaton field $S$ assuming that in this limit
the non-perturbative dynamics is dominated by the Yang-Mills gauge
forces. Nevertheless, the monodromy matrices we will obtain are not an
approximation in any sense, since the monodromy matrices are
of course field independent. They are just part of the full quantum
monodromy of the four-dimensional heterotic string.

Let us now precisely list the
assumptions we will impose when performing
the split of any of the semiclassical monodromies into the non-perturbative
ones:
\begin{enumerate}
\item
$\Gamma_{\infty}$ must be decomposed into precisely two factors.
\be
\Gamma_{\infty} = \Gamma_{M} \Gamma_{D}\label{twof}
\eq
\item
$\Gamma_{M}$ and therefore $\Gamma_{D}$ must be symplectic.
\item
$\Gamma_{M}$ must have a monopole like fixed point.
For the case of $w_1$, for instance, it must be of the form
\be
\left (N,-M \right) =
\left( 0,0,N^2, -N^2,0,0,0,0 \right)
\eq
\item
$\Gamma_{D}$ must have a dyonic fixed point.  For
the case of $w_1$, for instance, it must be of the form
\be
\left( N, -M \right) =
\left( 0,0,N^2, -N^2,0,0,-M_2, M_2 \right)
\eq
where $N^2$ and $M_2$ are proportional.
\item
$\Gamma_{M}$ and $\Gamma_{D}$ should be conjugated, that is, they must
be related by a change of basis, as it is the case in the rigid theory.
\item
The limit of large $S$ should be respected. This means that $S$
should only transform into a function of $T$ and $U$ (for at least
one of the four $SU(2)$ lines, as will be discussed in
the following).

\end{enumerate}

In the following we will show that under these assumptions the splitting
can be performed in a consistent way.  We will discuss the non perturbative
monodromies for the $SU(2)_{(1)}$ case in big detail.
Unlike
the rigid case, however, where
the decomposition of the perturbative monodromy into
a monopole like monodromy and a dyonic monodromy is unique (up to
conjugation), it will turn out that there are several distinct decompositions,
depending on four (discrete) parameters.
Only a subset
of these distinct decompositions should be,
however, the physically correct one.
One way of deciding which one is the physically correct one is to demand
that, when truncating this decomposition to the rigid case, one
recovers the rigid non perturbative monodromies of Seiberg/Witten.
This, however, requires one to have a reasonable prescription of
taking the flat limit, and one such prescription was given in section
(\ref{secseiwit}).

\subsection{Non-perturbative monodromies for $SU(2)_{(1)}$
\label{su21monopol}}

The non-perturbative part $f^{\rm NP}$ of the prepotential will depend on the
$S$-field.  We will make the following ansatz for the prepotential
\beqa
F =i \frac{X^1 X^2 X^3}{X^0} + (X^0)^2 \left(f(T,U) + f^{\rm NP} (S,T,U)
\right)
\eeqa
In order
to find a decomposition of
$\Gamma_{\infty}^{w_1}$, $\Gamma_{\infty}^{w_1}=
\Gamma_{M}^{w_1} \Gamma_{D}^{w_1}$,
we will now make the following ansatz:
$\Gamma_{\infty}^{w_1}$ has a peculiar block structure in that
the indices $j=0,1$ of the section $(P_j, i Q_j)$ are never
mixed with the indices $j=2,3$.
We will assume that $\Gamma_{M}^{w_1}$ and $\Gamma_{D}^{w_1}$ also have
this structure. This implies that the problem can be reduced
to two problems for 4 $\times$ 4 matrices.
Furthermore, we
will take $\Gamma_{M}^{w_1}$ to be the identity matrix
on its diagonal.
The existence of a basis
where the non--perturbative monodromies have this special form
will be aposteriori justified by the fact that it leads to
a consistent
truncation to the rigid case.

Taking into account all these assumptions yields
\cite{trunc} the following 8$\times$8
non--perturbative
monodromy matrices that depend on four parameters
$x,y,v,p$ and consistently describe the splitting of
the $T=U$ line
\be
\Gamma_{M}^{w_1} =
\left [\begin {array}{cccccccc} 1&0&0&0&0&0&0&0\\\noalign{\medskip}0&
1&0&0&0&0&0&0\\\noalign{\medskip}0&0&1&0&0&0&-2/3&2/3
\\\noalign{\medskip}0&0&0&1&0&0&2/3&-2/3\\\noalign{\medskip}x&y&0&0&
1&0&0&0\\\noalign{\medskip}y&v&0&0&0&1&0&0\\\noalign{\medskip}0&0&p&p
&0&0&1&0\\\noalign{\medskip}0&0&p&p&0&0&0&1\end {array}\right ]
\label{monopolew1}
\eq
$
\Gamma_D^{w_1}$ immediately follows from eq.(\ref{twof}).

The associated fixed points have the form
\be
\left(N,-M \right) =
\left( 0 , 0 , N^2, -N^2, 0, 0, 0, 0 \right)
\eq
for the monopole and
\be
\left(N, -M \right) =
\left(0, 0,  N^2, -N^2, 0, 0, \frac{3}{2}
N^2, -\frac {3}{2} N^2 \right)
\eq
for the dyon.

\subsection{Truncating the $SU(2)_{(1)}$ monopole monodromy
to the rigid case}

The monopole monodromy matrix for the first $SU(2)$, given in
equation (\ref{monopolew1}),
depends on 4 undetermined parameters, namely
$x,v,y$ and $p\neq 0$.
Note that
demanding the monopole monodromy matrix to be conjugated to the dyonic
monodromy matrix led to the requirement $p \neq 0$.
On the other hand, it follows from (\ref{monopolew1}) that
\beqa S \rightarrow S - i \left( y + v (TU - f^{\rm NP}_S) \right)
\label{dilmonow1}
\eeqa

Consider now the $4 \times 4$ monopole subblock which acts on $(P^2,P^3,
iQ_2,iQ_3)$
\beqa
\Gamma_{M23}^{w_1}=
\left [\begin {array}{cccc}
1&0&-2 \alpha &2 \alpha
\\\noalign{\medskip}0&1&2 \alpha &-2 \alpha
\\\noalign{\medskip}
p&p&1&0\\\noalign{\medskip}p&p&0&1\end {array}\right ]
\eeqa
($\alpha={1\over 3}$, $p\neq 0$.)
Rotating it into the Higgs basis gives that
\begin{equation}\eqalign{
\Gamma_{M}^{Higgs,w_1}&= M
\Gamma_{M23}^{w_1} M^{-1}\cr & =
\left [\begin {array}{cccc}
1&0&- 4\alpha \gamma^2&0
\\\noalign{\medskip}0&1&0&0
\\\noalign{\medskip}
0&0&1&0\\\noalign{\medskip}0&\frac{2p}{\gamma^2}&0&1\end {array}\right ] \cr}
\label{locmonohiggs}
\end{equation}
where $\alpha={1\over 3}$, $p\neq 0$ and
$M$ is given in equation (\ref{higgsbasis}).
In the rigid case, on the other hand, one expects to find
for the rigid monopole monodromy matrix in the Higgs basis that
\beqa
\tilde{\Gamma}_{M}^{Higgs,w_1}=
\left [\begin {array}{cccc}
1&0&- 4\tilde{\alpha}\gamma^2&0
\\\noalign{\medskip}0&1&0&0
\\\noalign{\medskip}
0&0&1&0\\\noalign{\medskip}0&\frac{2\tilde{p}}{\gamma^2}
&0&1\end {array}\right ]
\label{rigidmono}
\eeqa
$(\tilde\alpha={1\over 4}$, $\tilde p=0$.)
The first and third lines of (\ref{rigidmono}) are,
for $\tilde{\alpha}=\frac{1}{4}$, nothing but the monodromy matrix
for one $SU(2)$ monopole ($\gamma^2 = 2$ in the conventions of
Seiberg/Witten \cite{SW1},
and $\gamma^2 = 1 $ in the conventions of Klemm et al \cite{KLT}).

Thus, truncating the monopole monodromy matrix (\ref{monopolew1})
to the rigid case appears to produce jumps in the parameters $p \rightarrow
\tilde{p}=0$ and $\alpha \rightarrow \tilde{\alpha}$ as given above.
In \cite{trunc}  we  presented a field theoretical explanation
for the jumps occuring in the parameters $p$ and $\alpha$ when
taking the rigid limit.
This explanation also determines, as a
bonus,  the values of the
parameters $v,y$ and $p$: $y={8\over 3}$, $p={4\over 3}$, $v=0$.
Moreover, one can show that, in order
to decouple the four $U(1)$'s at the non-perturbative level, one
has to have $x=v$ and consequently $x=0$.
Note that $v=0$
ensures that $S \rightarrow S - i y$ under the $SU(2)_{(1)}$
monopole monodromy.

\section{Conclusions}

In this paper we have first discussed the properties of the
perturbative prepotential of $N=2$ heterotic strings. The
derivatives of the holomorphic prepotential determine the
low energy effective Lagrangian
of the $N=2$ heterotic strings, such as the effective gauge couplings or
the K\"ahler potential. At the one-loop level the effective couplings
are related to automorphic functions in a very interesting way.
In addition we
have shown
that the semiclassical monodromies associated with lines of enhanced
gauge symmetries can be consistently split into pairs of non-perturbative
lines of massless monopoles and dyons. Furthermore, all monodromies
obtained in the string context allow for a consistent truncation to
the rigid monodromies of \cite{SW1,KLTY,KLT}.
Recently the non-perturbative monodromies for the  models
with the two fields $S$ and
$T$
and their rigid limits were also determined by using
the string-string duality between heterotic strings and four-dimensional
type II strings, compactified on a suitably chosen Calabi-Yau  space.
The perturbative monodromy (in
this case $T \rightarrow \frac{1}{T}$) and its decomposition into
non-perturbative monopole and dyon monodromies, as computed from
the type II Calabi-Yau side, agree with our perturbative and
non-perturbative monodromies
after introducing a compensating
shift for the dilaton, $S \rightarrow S - i$, which is generated
by $\Gamma_S$ in (\ref{sdualpert}). We will come back to these
issues in more detail in a future paper \cite{comp}.

\end{document}